\documentclass[fleqn,10pt]{wlscirep}
\usepackage[utf8]{inputenc}
\usepackage[T1]{fontenc}
\usepackage{xcolor}

\title{The mixing-demixing phase diagram of ultracold heteronuclear mixtures in a ring trimer}

\author[1,*]{Andrea~Richaud}
\author[2]{Alessandro~Zenesini}
\author[1]{Vittorio~Penna}
\affil[1]{Department of Applied Science and Technology and u.d.r. CNISM, Politecnico di Torino, I-10129 Torino, Italy}
\affil[2]{Institut f\"ur Quantenoptik, Leibniz Universit\"at Hannover, 30167~Hannover, Germany}

\affil[*]{andrea.richaud@polito.it}

\begin{abstract}
We derive the complete mixing-demixing phase-diagram relevant to a bosonic binary mixture confined in a ring trimer  and modeled within the Bose-Hubbard picture. The mixing properties of the two quantum fluids, which are shown to be strongly affected by the fragmented character of the confining potential, are evaluated by means of a specific indicator imported from Statistical Thermodynamics and are shown to depend only on two effective parameters incorporating the asymmetry between the heteronuclear species. To closely match realistic experimental conditions,  our study is extended also beyond the pointlike approximation of potential wells by describing the systems in terms of two coupled Gross-Pitaevskii equations. The resulting mean-field analysis confirms the rich scenario of mixing-demixing transitions of the mixture and also constitutes an effective springboard towards a viable experimental realization. We additionally propose an experimental realization based on a realistic optical-tweezers system and on mixtures of bosonic $^{23}\mathrm{Na}$ and $^{39}\mathrm{K}$, thanks to the large tunability of their intra- and inter-species scattering lengths.   
\end{abstract}
\begin{document}

\flushbottom
\maketitle

\thispagestyle{empty}

\section*{Introduction}
Ultracold mixtures of different atomic species have a long experimental history both for studies in bulk \cite{PhysRevLett.99.010403} and in optical lattices \cite{Gadway,Soltan,PhysRevA.77.011603,Inguscio} and allowed the investigation of a plethora of intriguing phenomena such as entropy exchange\cite{PhysRevLett.103.140401}, miscibility in degenerate gases\cite{PhysRevA.94.013602}, beyond-mean-field effects\cite{PhysRevLett.115.155302, Cabrera301, PhysRevLett.120.235301}, universality of three-body losses\cite{PhysRevLett.111.105301, PhysRevLett.112.250404,PhysRevLett.117.163201} and ultracold chemistry \cite{ChemicalReviews}.

Fragmentation of bosonic mixtures in arrays of potential wells constitutes an important factor in determining their mixing properties and plays a crucial role, together with the repulsion between the mixture's components, in triggering spatial phase separation. A large amount of recent theoretical studies have revealed a colorful phenomenology ranging from phase separation mechanism \cite{sep1,sep3}, demixing of dipolar mixtures\cite{sep2}, and quantum emulsions \cite{Buonsante_PRL,qe1} to the critical properties of quasiparticles spectrum across the transition\cite{Angom}, thermal suppression\cite{Roy} and the role of entanglement\cite{NoiEntropy} in phase separation, and cascades of quantum phase transitions of vector bosons\cite{Belemuk}.


Recently, phase separation has been investigated for a binary mixture in a two-well trap (dimer). This constitutes the simplest geometry whose fragmented character exacerbates the competition between \textit{intra}- and \textit{inter}-species repulsions\cite{Bruno} therefore affecting the occurrence of spatial separation. In this system, in fact, the transition from a mixed to a fully demixed ground-state, which transparently emerges within the Continuous Variable Picture (CVP)\cite{PennaLinguaPRE}, is confirmed by the analysis of genuinely quantum indicators, such as the energy spectrum, the entanglement between the species, and the quantum tomography of the ground state \cite{PennaLinguaJPB,NoiEntropy}. The study of phase-separation mechanism has been extended to the ring-trimer geometry in the simple case of twin species \cite{NoiSREP}, i.e. species featuring the same hopping amplitude $T$, the same on-site interaction $U$, and the same number of atoms $N$. Despite its simplicity, this toy model has already shown a non-trivial phenomenology, marked by the presence of an unexpected intermediate phase which is in between a perfectly mixed and a fully demixed phase. 

In this work, we aim at ferrying this intriguing but maybe too simplistic model towards an actual experimental realization. In doing so, we take into account the presence of asymmetries $T_a\neq T_b$, $U_a\neq U_b$ and $N_a\neq N_b$ between atomic species \textit{a} and \textit{b}, a circumstance which makes the model we are dealing with more realistic. The increased number of variables that come into play therefore makes the developed analysis remarkably more articulated and comprehensive than the original one, but significantly broadens its scope. Such study results in a rather rich two-dimensional phase diagram, which includes four different phases, each one marked by a specific boson redistribution among the wells and featuring a different degree of mixing. In order to corroborate our previsions, we extend our analysis beyond the pointlike-site approximation (reducing the local potential traps to dimensionless sites) which intrinsically characterizes both the Bose-Hubbard model and the CVP. To this end, we present a mean-field (MF) approach based on the solution of two coupled stationary Gross-Pitaevskii equations (GPEs). This technique, which allows one to take into consideration the boson distributions not only \textit{among} but also \textit{within} the wells, lends itself to direct interlink with a real-world experiment and therefore represents an effective bridge between the quite abstract (but rather clarifying) CVP model and a real physical system. 

\section*{The model}
In this paper, we consider a bosonic binary mixture confined in a three-well potential with periodic boundary conditions (i.e. a ring trimer). This is well described by the Bose-Hubbard Hamiltonian
\begin{equation}
\label{eq:BH}
		 H= - T_a \sum_{j=1}^{3} \left(a_{j+1}^\dagger a_j +a_j^\dagger a_{j+1} \right) + \frac{U_a}{2} \sum_{j=1}^{3} n_j(n_j-1) - T_b \sum_{j=1}^{3} \left(b_{j+1}^\dagger b_j +b_j^\dagger b_{j+1} \right)+\frac{U_b}{2} \sum_{j=1}^{3} m_j(m_j-1)+W \sum_{j=1}^{3} n_j\, m_j, 
\end{equation}
a second-quantized model where operator $a_i$ ($a_i^\dagger$) destroys (creates) a species-a boson in site $i\in\{1,2,3\}$ (with $i=4\equiv 1$ due to the ring geometry). The same holds for operators $b_i$ and $b_i^\dagger$, which destroy and create species-b bosons. The bosonic character of these operators is enforced by standard commutation relations, i.e $[a_i,a_\ell^\dagger]=\delta_{i,\ell}=[b_i,b_\ell^\dagger]$, $[a_i,b_\ell^\dagger]=0$. After introducing number operators $n_i=a_i^\dagger a_i$ and $m_i=b_i^\dagger b_i$, one can observe that the total numbers of bosons of each species ($N_a=\sum_i n_i$ and $N_b=\sum_i m_i$ respectively) constitute two independent conserved quantities, namely $[N_a,H]=[N_b,H]=0$. As regards model parameters, $T_a$ and $T_b$ are hopping amplitudes, $U_a$ and $U_b$ represent intra-species repulsive interactions, and $W$ represents the inter-species repulsion.  

\paragraph{Semiclassical approach towards demixing.} The CVP, originally introduced to investigate the spatial fragmentation of a condensate in a two-well potential \cite{Spekkens}, is a semi-classical approximation scheme based on the replacement of the inherently discrete quantum numbers associated to the Fock-state basis with continuous variables. This technique has proved to be particularly effective in capturing the essential critical behaviour of complex many-body systems \cite{Penna_Burioni,Ciobanu,Zin}. It also allows one to derive an effective semi-classical Hamiltonian $H_{eff}$ which well reproduces the low-energy physics of the original quantum model \cite{Javanainen, Dynamical_Bifurcation, Mazzarella_Penna} and to turn the search for the ground state thereof into that of the minimum of the potential $V$ provided by $H_{eff}$. In the current case, the CVP associates to Hamiltonian (\ref{eq:BH}) the potential
\begin{equation}
\label{eq:V}
\begin{split}
V=  &-2N_a T_a\left(\sqrt{x_1x_2}+\sqrt{x_2x_3}+\sqrt{x_3x_1}\right) -2N_b T_b \left(\sqrt{y_1y_2}+\sqrt{y_2y_3}+\sqrt{y_3y_1}\right)+ \\
 & +\frac{U_aN_a^2}{2}\left(x_1^2+x_2^2+x_3^2\right) + \frac{U_bN_b^2}{2}\left(y_1^2+y_2^2+y_3^2 \right)+ WN_a N_b \left(x_1y_1+x_2y_2+x_3 y_3 \right)
  \end{split}
\end{equation}
which comes with the two constraints $\sum_i x_i=\sum_i y_i =1$, enforcing particle number conservation in both condensed species. Notice that variables $x_i:=n_i/N_a$ and $y_i:=m_i/N_b$ represent normalized boson populations and are regarded as \textit{continuous} in view of the fact that the total numbers of bosons $N_a$ and $N_b$ are assumed to be large. 
 
\section*{Results}
We investigate the ground-state properties of a bosonic binary mixture in a ring trimer, showing that they can be conveniently represented by means of a two-dimensional phase diagram which is, in turn, spanned by two specific effective variables representing functions of the original-model parameters. Each phase features a characteristic boson-population distribution, which results in a different degree of mixing and in a different expression of the ground-state energy.   

\begin{figure}[h!]
\centering
\includegraphics[width=.8\linewidth]{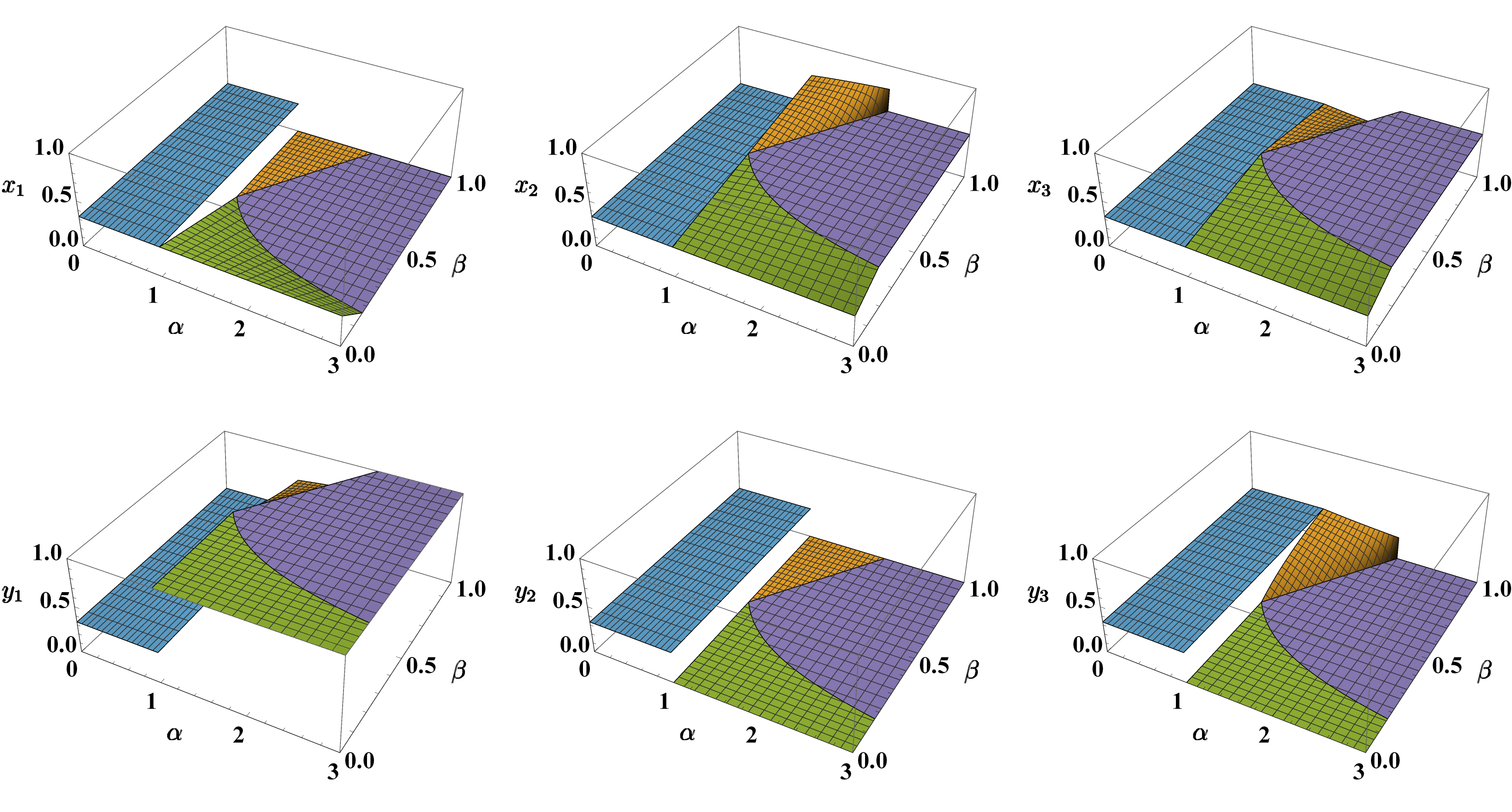}
\caption{Summary of the configurations $(\vec{x},\vec{y})$ which minimize effective potential (\ref{eq:V_*}) (under the constraints $\sum_j x_j = \sum_j y_j =1$) in the four phases. Phase 1 (in blue) features perfect demixing; phase 2 (in orange) features demixing in two out of three sites and mixing in the third one; phase 3 (in purple) is the fully-demixed one; phase 4 (in green) is the one where species $b$ clots in a site while species $a$ spreads in all three sites, but in different proportions. }
\label{fig:Popolazioni_T_0}
\end{figure}

\paragraph*{The phase diagram.}
The presence of four different phases emerges in a crystal-clear way when one considers the large-populations limit (this can be seen as sort of thermodynamic limit according to the scheme highlighted in \cite{Dynamical_Bifurcation}), i.e. when ratios $U_a N_a/T_a$ and $U_b N_b/T_b$ are big enough to make hopping terms negligible. In this limit, in fact, the (rescaled version of) effective potential (\ref{eq:V}), namely $V/(U_aN_a^2)$, reduces to 
\begin{equation}
\label{eq:V_*}
  V_*= \frac{1}{2}\left(x_1^2+x_2^2+x_3^2\right) + \frac{\beta^2}{2}\left(y_1^2+y_2^2+y_3^2 \right)+ \alpha\beta \left(x_1y_1+x_2y_2+x_3 y_3 \right),
\end{equation}
an expression including only \textit{two} effective parameters, 
\begin{equation}
\label{eq:alfa_beta}
\alpha = \frac{W}{\sqrt{U_a U_b}}, \qquad \beta=\frac{N_b}{N_a} \sqrt{\frac{U_b}{U_a}}.
\end{equation} 
The former corresponds to the ratio between the inter-species and the (geometric average of) the intra-species repulsive interactions while the latter accounts for the degree of asymmetry between the two species. This reduced set of parameters proves to be the most natural one to investigate the occurrence of mixing-demixing transitions in the system and is therefore employed to build the phase diagram.       
\begin{figure}[h!]
\centering
\includegraphics[width=0.4\linewidth]{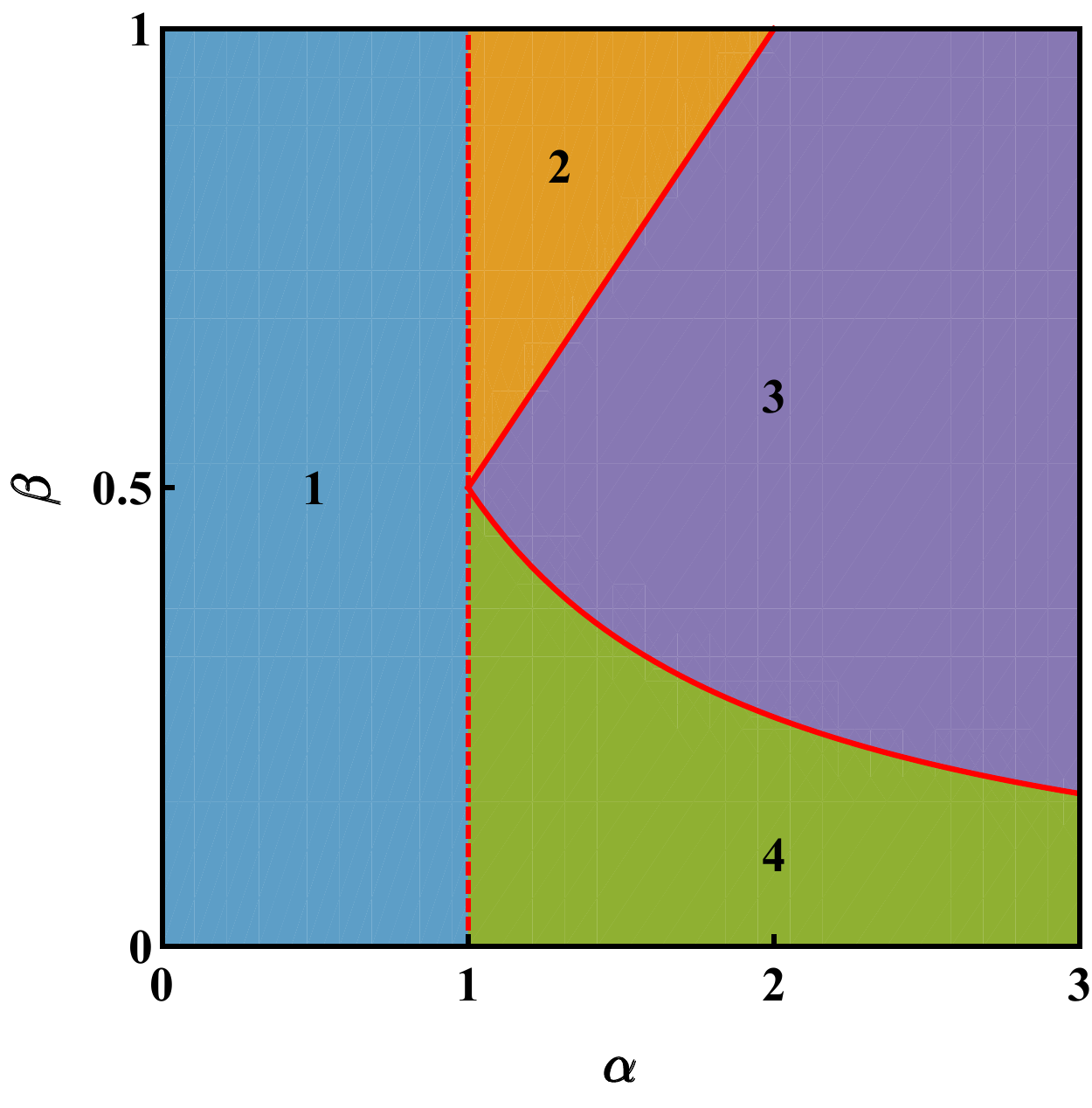}
\caption{Phase diagram of a binary mixture in a ring trimer. Each of the four phases corresponds to a different functional relationship between the minimum-energy configuration and effective model parameters (\ref{eq:alfa_beta}). This circumstance has a direct impact on the mixing properties of the system and on its ground-state energy. Red dashed (solid) line represents a mixing-demixing transition across which the components $x_j$ and $y_j$ of the minimum-energy configuration feature a jump discontinuity (are continuous). The plot has been obtained by means of a fully-analytic minimization of potential (\ref{eq:V_*}) under the constraints $\sum_j x_j =\sum_j y_j =1$ .}
\label{fig:Diagramma_di_fase}
\end{figure}
In this regard, the region $\mathcal{D}=\{(\alpha,\beta)\,:\, 0\le\alpha\le 3, \, 0\le\beta\le 1 \}$, which is depicted in Fig. \ref{fig:Diagramma_di_fase} already includes all four phases that the system's ground state can exhibit (notice, in fact, that if $\beta>1$, one can swap species labels $a$ and $b$ and therefore come back to region $\mathcal{D}$). In each phase, the configuration $(\vec{x},\vec{y})$ which minimizes effective potential (\ref{eq:V_*}) is marked by a specific dependence on the effective model parameters (\ref{eq:alfa_beta}). Analytic expressions $x_j(\alpha,\beta)$ and $y_j(\alpha,\beta)$ have been derived by means of an exhaustive exploration of the polytope-like domain of function $(\ref{eq:V_*})$ (see \cite{NoiSREP} for a detailed description of the method), and are presented here, together with the essential features of the associated phase:

\begin{enumerate}
\item  In phase 1, being $\alpha<1$, the inter-species repulsion is too small to trigger spatial phase separation and the two species are uniformly distributed among the three wells, i.e. $x_j=y_j=1/3, \, \forall j=1,2,3$.

\item  Phase 2 occurs for $1<\alpha<2\beta$, that means for intermediate values of $\alpha$ and not too asymmetric species. In this region, one has complete demixing in two wells and mixing in the remaining one. The explicit expression of the minimum-energy configuration as functions of model parameters (\ref{eq:alfa_beta}) reads
$$
       x_1= 0, \qquad x_2= \frac{\alpha^2-\alpha \beta-2}{\alpha^2-4}, \qquad x_3= \frac{\alpha \beta-2}{\alpha^2-4},
$$
$$
     y_1= \frac{\alpha^2\beta -\alpha -2 \beta}{(\alpha^2-4)\beta}, \qquad y_2=0,  \qquad y_3= \frac{\alpha-2 \beta}{\left(\alpha^2-4\right) \beta}.
$$

\item Phase 3, occurring for sufficiently high values of $\alpha$ and not too asymmetric species (more specifically, for $1/(2\alpha)<\beta<\alpha/2$), features a completely demixed scenario, such that species $b$ clots in one well, while species $a$ equally spreads in the other two sites: 
$$
       x_1 = 0, \quad x_2 = \frac{1}{2}, \quad x_3=\frac{1}{2},
 $$
 $$
       y_1 = 1, \quad y_2 = 0, \quad y_3=0.
 $$
\item Phase 4 occurs for sufficiently high values of $\alpha$ and sufficiently asymmetric species (i.e. for $\alpha>1$ and $\beta<1/(2\alpha)$). A good degree of asymmetry can be achieved, for example, if species $b$ constitutes an impurity with respect to the majoritarian species $a$, meaning that $N_b\ll N_a$. The hallmark of this phase is that species $b$  clots in one site, while species $a$ spreads in all three sites, but in different proportions:  
$$
       x_1 = \frac{1}{3} (1-2 \alpha \beta), \quad x_2 =\frac{1}{3} (1 + \alpha \beta), \quad x_3=\frac{1}{3} (1+ \alpha \beta),
 $$
 $$
       y_1 = 1, \quad y_2 = 0, \quad y_3=0.
 $$

\end{enumerate}
An illustrative summary of all presented minimum-energy configurations is provided in Fig. \ref{fig:Popolazioni_T_0}. Concerning the ones discussed at points 2, 3 and 4, we remark that, due to the $Z_3$ symmetry of the system, they are not unique and that other isoenergetic configurations can be obtained by cyclic permutations of site indexes.
Quantum-mechanically, the ground state is degenerate only in the infinite-population limit because, as soon as they come into play, hopping terms lift the degeneracy and the ground state's structure gets that of a Schr\"{o}dinger cat \cite{Ciobanu,Wang_Penna,NoiSREP,NoiEntropy}. For example, in phase 3, the ground-state is of the type $|E_0 \rangle\approx \frac{1}{\sqrt{3}}\left(|0,\frac{1}{2},\frac{1}{2}\rangle_a |1,0,0\rangle_b  + |\frac{1}{2},0,\frac{1}{2}\rangle_a |0,1,0\rangle_b  +  |\frac{1}{2},\frac{1}{2},0\rangle_a |0,0,1\rangle_b	\right) $, an expression where symbol ``$\approx$" reminds that minor contributions coming from Fock states with different boson distributions and activated by the non-zero hopping amplitudes have been understood.    

As clearly illustrated in Fig. \ref{fig:Popolazioni_T_0}, for $\beta < 1$, the minimum-energy configuration $(\vec{x},\vec{y})$, as a function of model parameters $\alpha$ and $\beta$, is discontinuous at transitions 1-2 and 1-4 while it is continuous at transitions 2-3 and 3-4.  Nevertheless, in the case of perfectly symmetric species, $\beta = 1$, the system gains an additional symmetry (consisting in the interchangeability of species' labels in formulas (\ref{eq:BH}) - (\ref{eq:V_*})) and a further discontinuity at transition 2-3 appears. The twin-species limit, widely discussed in \cite{NoiSREP}, therefore features a qualitatively different critical behavior.

\begin{figure}[h!]
\centering
\includegraphics[width=0.4\linewidth]{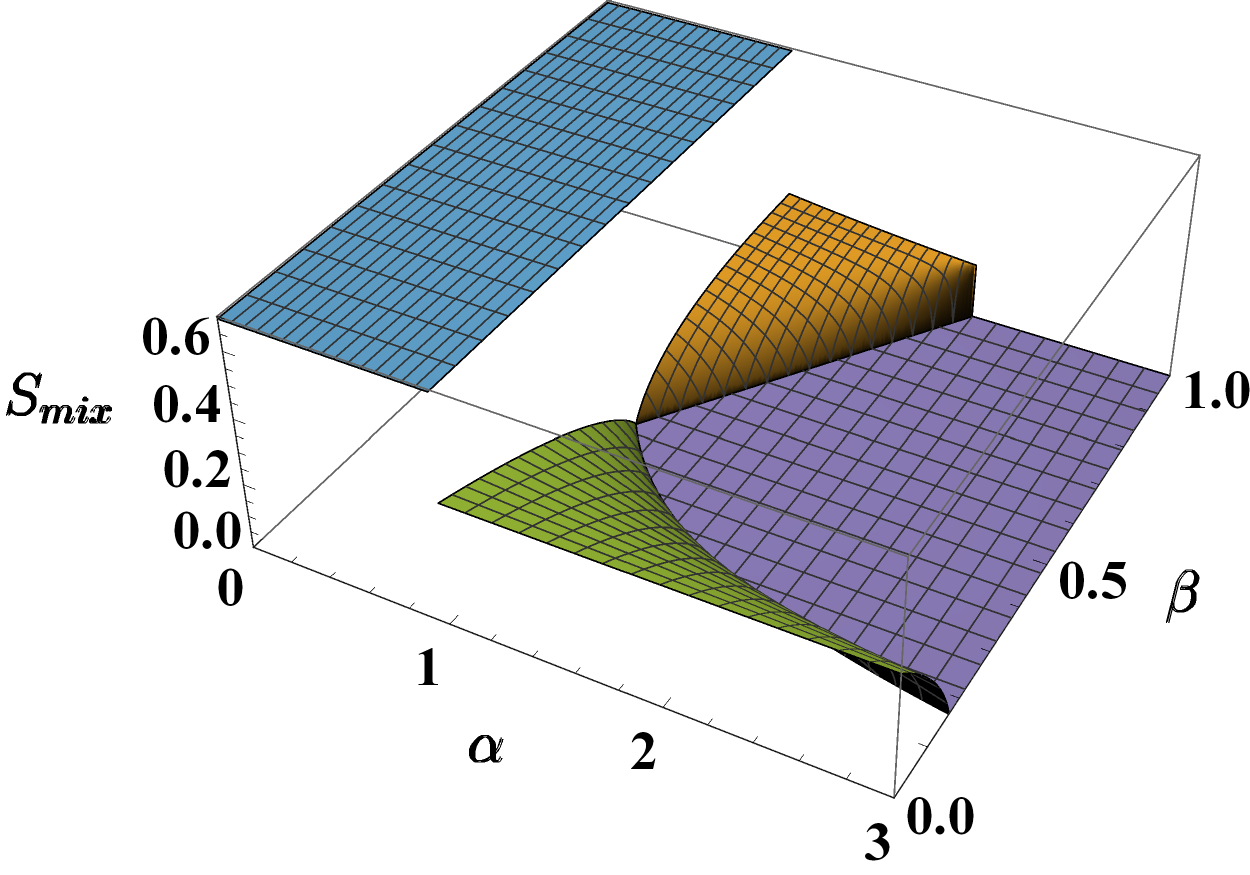}
\caption{Entropy of mixing, $S_{mix}$, in the four phases. Phase 1 (in blue) features perfect mixing and $S_{mix}$ takes the biggest possible value, i.e. $\log 2\approx 0.69$. Conversely, phase 3 (in purple) features perfect demixing and $S_{mix}$ is therefore zero. }
\label{fig:Entropy_T_0}
\end{figure}

\paragraph*{Phases and degree of mixing.}
An effective indicator to quantify the degree of mixing of two different species in discretized domains is the entropy of mixing ($S_{mix}$). Originally introduced in the context of macromolecular simulations\cite{Camesasca}, this measure has been recently introduced in the realm of ultracold atoms in order to investigate the link between chaotic dynamical regimes and mixing properties of a bosonic binary mixture in a ring trimer \cite{Noi_NJP}. According to the definition given in \cite{Camesasca}, the entropy of mixing associated to a certain minimum-energy configuration $(\vec{x},\vec{y})$ reads 
\begin{equation}
\label{eq:S_mix}
  S_{mix} = -\frac{1}{2}\sum_{j=1}^3 \left( x_j \log\frac{x_j}{x_j+y_j} +  y_j \log\frac{y_j}{x_j+y_j} \right).
\end{equation}
In comparison to the one used in \cite{Brandani,Noi_NJP}, this formula is based on \textit{normalized} populations $x_j$ and $y_j$ (rather than $n_j$ and $m_j$) and is therefore more suited to describe binary mixtures where the species feature a particle-number imbalance (moreover the contribution of each site $j$ to the total entropy of mixing is not fixed, but it is weighted by the fractions of particles present therein). As shown in Fig. \ref{fig:Entropy_T_0}, $S_{mix}$ is zero in phase 3 (perfect demixing) while achieves the maximum possible value, $\log 2 \approx 0.69$ in phase 1 (perfect mixing). More generally, $S_{mix}$, as a function of model parameters $\alpha$ and $\beta$, mirrors the criticalities exhibited by the minimum-energy configuration, i.e. it is discontinuous at transition 1-2 and 1-3 while it is continuous at transitions 2-3 (apart from the special case $\beta=1$) and 3-4. For this reason, it constitutes a valid indicator to capture the occurrence of mixing-demixing transitions. Note that $S_{mix}$ is particularly advantageous in that it allows one to represent the system critical behavior by avoiding its description in terms of six boson populations $x_j$ and $y_j$.

\begin{figure}[h!]
\centering
\includegraphics[width=.8\linewidth]{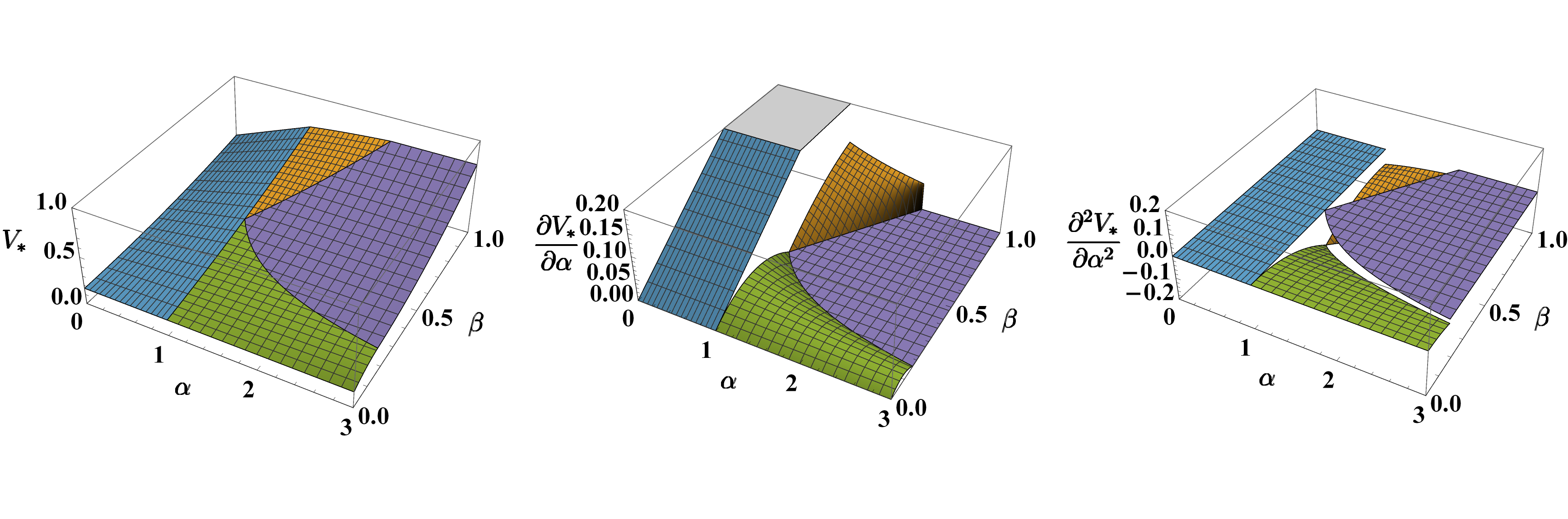}
\caption{Effective potential (\ref{eq:V_*}) (first panel) and its first and second derivative with respect to $\alpha$ (second and third panel) relevant to the minimum-energy configuration as functions of model parameters parameters $\alpha$ and $\beta$.}
\label{fig:Effective_potential}
\end{figure}

\paragraph*{Phases and free energy.}
Being the temperature zero, the free energy $F=E-TS$ coincides with the internal energy $E$, i.e. the ground-state energy. The latter, within the CVP approach, can be computed by means of the effective potential (\ref{eq:V_*}) which, for the four boson distributions corresponding to the four phases,  exhibits a specific dependence law on model parameters $\alpha$ and $\beta$ given by
$$
	V_{*,1}=\frac{1}{6} \left(2 \alpha \beta+\beta^2+1\right), \qquad 
    V_{*,2}=\frac{\left(\alpha^2-2\right) \beta^2+\alpha^2-2 \alpha \beta-2}{2 \left(\alpha^2-4\right)},
$$
$$
    V_{*,3}=\frac{1}{4} \left(2 \beta^2+1\right), \qquad 
    V_{*,4}=\frac{1}{6} \left[\beta \left(-2 \alpha^2 \beta+2 \alpha+3 \beta\right)+1\right].
$$
The graphic representation of these expressions (see first panel of Fig. \ref{fig:Effective_potential}), shows that $V_*$ is indeed continuous everywhere and, in particular, across the transitions. Nevertheless, in agreement with the previous mixing-entropy analysis, the non-analytic character of the ground-state energy does emerge if one computes the first and the second derivative of $V_*$ with respect to $\alpha$, which is regarded as a control parameter. As depicted in the second and in the third panel of the aforementioned figure, in fact, at transitions 1-2 and 1-4, the first derivative is discontinuous, while at transitions 2-3 and 3-4, the first derivative is continuous but the second one is discontinuous (except for the special case $\beta=1$, where the first derivative is discontinuous at transition 2-3). 

\subsection*{Finite-size effects on the mixing-demixing transitions}
\begin{figure}[h!]
\centering
\includegraphics[width=1\linewidth]{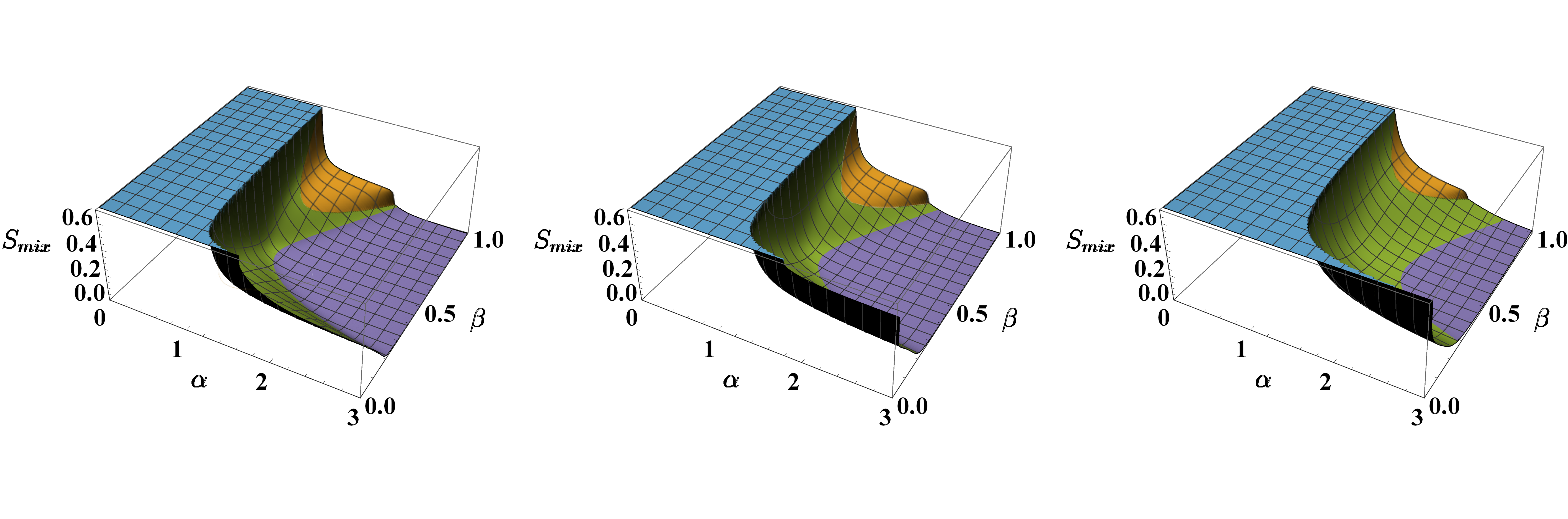}
\caption{ Entropy of mixing associated to the configurations $(\vec{x},\vec{y})$ which minimize effective potential (\ref{eq:V}) sweeping model parameters $\alpha$ and $\beta$. The following values/intervals have been chosen: $U_a=U_b=1$, $N_a=15$, $N_b\in [0,15] \, \Rightarrow\, \beta \in [0,1]$, $W\in[0,3]\,\Rightarrow\,\alpha\in[0,3]$, $T_a=T_b=T$, where $T=0.1,\,0.2,\,0.5$ in the left, central and right panel respectively. Colors have been employed as a guide to the eye: blue is used when $S_{mix}=\log 2$, purple when $S_{mix}<0.05$, green for all intermediate values except the dome corresponding to phase 2 and therefore colored in orange.}
\label{fig:Entropy}
\end{figure}

As already mentioned, the four different phases emerge at their clearest only in the large-populations limit, i.e. when $U_a N_a/T_a\to \infty$ and $U_b N_b/T_b\to \infty$. Such phases are still recognizable also in the more realistic case of finite-size systems (i.e. systems with limited numbers of atoms and featuring non-vanishing hopping terms), although the phase diagram presented in Fig. \ref{fig:Diagramma_di_fase} gets blurred and deformed. The effects of finite ratios $U_a N_a/T_a$ and $U_b N_b/T_b$ can be summarized as follows:
\begin{itemize}
	\item The minimum-energy configuration $(\vec{x},\vec{y})$, regarded as a function of model parameters $\alpha \in (0,3)$ and $\beta \in (0,1)$, is continuous and, as a consequence, so is the entropy $S_{mix}$ (comparing Fig. \ref{fig:Entropy_T_0} and Fig. \ref{fig:Entropy}, one can notice that the jump discontinuities give way to smooth junctions). 
    \item The fully-mixed phase is favored by the presence of non-negligible hopping amplitudes, its border being given by the inequality
\begin{equation}
\label{eq:Inequality}
    \alpha< \sqrt{\left(1 +\frac{9}{2} \frac{T_a}{U_a N_a}\right)\left(1 +\frac{9}{2}\frac{T_b}{U_b N_b}\right)}.
\end{equation}
The latter represents the condition under which the Hessian matrix associated to effective potential (\ref{eq:V}) (and evaluated at point $x_j=y_j=1/3$, with $j=1,\, 2,\, 3$) is positive definite. Walking away from the large-populations limit, the right-hand term rises above the value $1$, thus determining an enlargement of phase 1 at the expenses of the neighboring phases (see the enlargement of the blue region in Fig. \ref{fig:Entropy} moving from the left panel to the right panel). It is worth mentioning that (the twin-species limit of) condition (\ref{eq:Inequality}) was found to define the region of parameters' space where the spectrum of Bogoliubov quasiparticles is well-defined and not degenerate\cite{NoiPRA2}.  
    \item As intuition suggests, the fully demixed phase is mined by the presence of hopping processes between the wells and therefore occurs for higher values of $\alpha$ (see the shrinking of the purple region in Fig. \ref{fig:Entropy} moving from the left panel to the right panel). This clearly reminds the physics of the Superfluid-to-Mott-insulator transition \cite{Greiner2002}, where a stronger on-site interaction is required to observe the transition when a larger hopping rate is present.
    \item Interestingly, increasing the hopping amplitudes, phase 4 not only enlarges, but invades the $\beta>1/2$ region, crushing and shrinking phases 2 and 3 (moving from the left to the right panel of Fig. \ref{fig:Entropy}, one can see that the green region enlarges at the expenses of the orange and of the purple regions).   
\end{itemize}

\section*{Toward an experimental realization: a "real" mixture and  beyond point-like potential wells}
In the previous section we have discussed the mixing properties of an asymmetric binary mixture within the CVP of the BH model. In order to get closer to an experimental realization, one has to consider also the non-pointlike character of potential wells and the actual tunability of the scattering lengths in atomic systems.

\begin{figure}[h!]
\centering
\includegraphics[width=.4\linewidth]{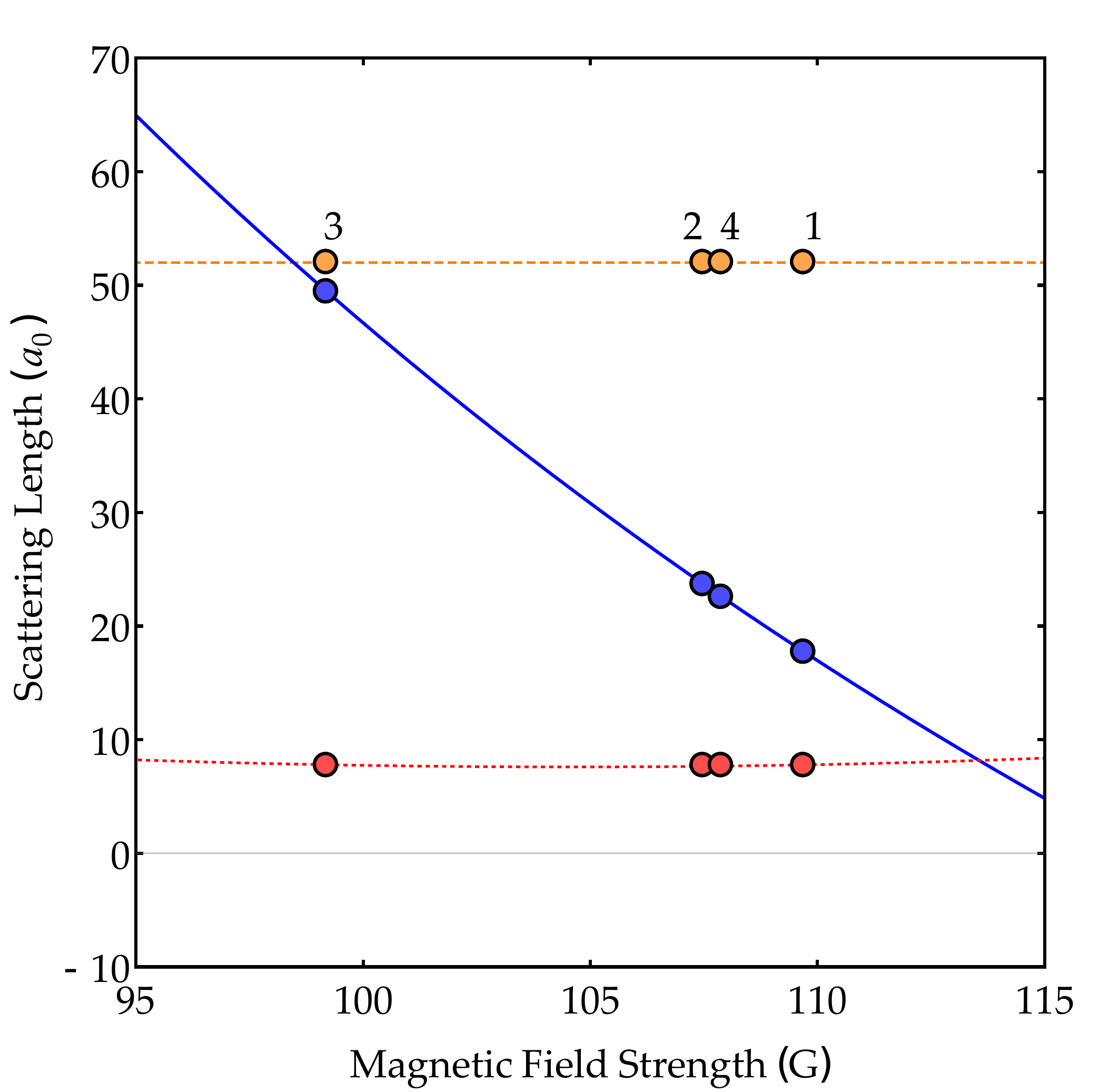}
\includegraphics[width=0.37\linewidth]{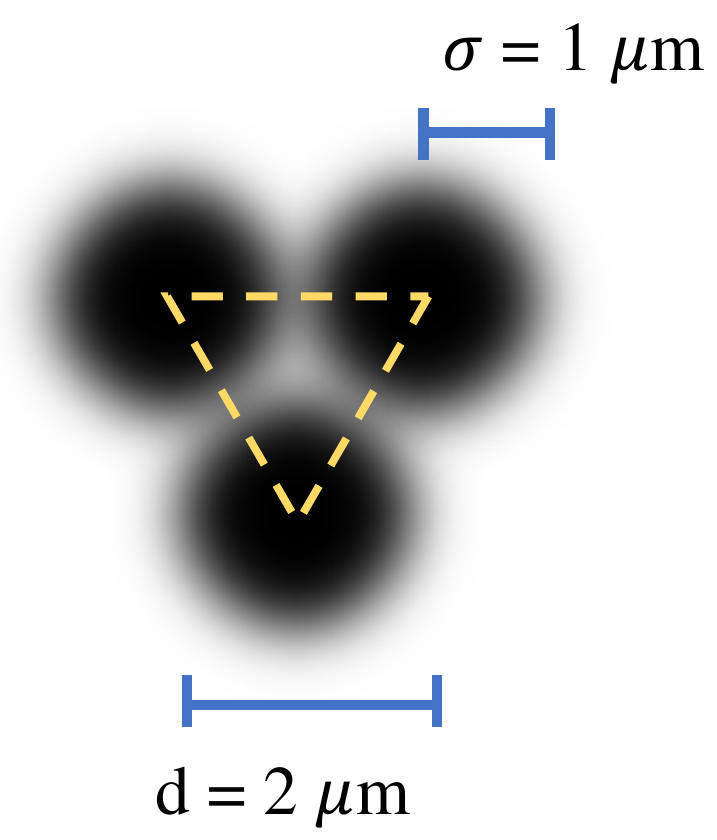}
\caption{Left panel: Intra- and inter-species scattering lengths for $|f=1,m_f=-1\rangle_{\mathrm{^{23}Na}}+|f=1,m_f=-1\rangle_{\mathrm{^{39}K}}$ mixture given in the Bohr radii $a_0$. Yellow dashed, red dotted and blue continuous line refer to sodium $a_{\mathrm{Na}}$, potassium $a_{\mathrm{K}}$ and sodium-potassium $a_{\mathrm{Na,K}}$ scattering lengths. Labels refer to points considered in the next section. Note the approximately constant intra-species scattering lengths and the inter-species  $a_{\mathrm{Na,K}}$ varying from < $a_{\mathrm{K}}$ to > $a_{\mathrm{Na}}$. Right panel: Schematic representation of the three gaussian traps constituting the ring-trimer potential. Yellow dashed line represents the equilateral triangle formed by the trap's center. Such triangle can be associated to the curvilinear abscissa $x$ present in equations (\ref{eq:Eig_prob_1}) and (\ref{eq:Eig_prob_2}) (and therefore automatically endowed with periodic boundary conditions).}
\label{fig:Scattering}
\label{fig:Potenziale_ottico}
\end{figure}

Despite the large tunability of the scattering length via magnetic Feshbach resonances \cite{RevModPhys.82.1225}, this knob does not allow for an independent tuning of the intra- and the inter-species scattering lengths. Some mixtures have extremely large background scattering lengths \cite{PhysRevA.79.042718}, very narrow resonances far beyond experimental magnetic field stability \cite{PhysRevA.79.042718}, overlapping resonances \cite{PhysRevA.97.023623}, and promising broad inter-species resonances where, however, the inter-species scattering length is excessively large or has the wrong sign \cite{PhysRevA.94.013602}. 

In this prospective, $^{23}\mathrm{Na} + ^{39}\mathrm{K}$ constitutes an ideal mixture. The spin mixture $|f=1,m_f=-1\rangle_{Na}+|f=1,m_f=-1\rangle_{K}$, where $f$ is the atomic total angular momentum and $m_f$ its projection on the quantization axis, has been recently doubly Bose-condensed and presents a favorable window of magnetic field (from 90 to 150\,G) where the combination of two intra- and two inter-species resonances allow for a smooth tuning of the scattering lengths \cite{PhysRevA.97.023623} (see left panel of Fig.\,\ref{fig:Scattering}). Additionally, in this range of magnetic field, three-body losses \cite{PhysRevLett.91.123201} are limited thanks to the small value of $a_{\mathrm{Na,K}}$. In our case, a change of the magnetic field corresponds to a shift along a line in the phase diagram of Fig.\,\ref{fig:Funzionali_energie}, where $\alpha$ and $\beta$ (horizontal and vertical axis respectively) have been rewritten in terms of MF parameters. In the region of interest, the scattering lengths of sodium and potassium are almost constant and the line will be almost horizontal, crossing the different phases at a value $\beta$ mainly determined by the atom-number ratio.

From now on, the formulas will be strictly linked with the real experimental setup and labels $a$ and $b$ will stand for \textit{sodium} and \textit{potassium} respectively. Parameters $m_a=22.99$ u and $m_b=39.10$ u are therefore the atomic masses of such elements, where u stands for unified atomic mass unit.

Recent developments in single-site resolution detection of single atoms and the realization of diffraction limited potentials have pushed forward the capabilities of handling few atoms in tweezers systems \cite{PhysRevX.2.041014}. The trimer potential consists (see right panel of Fig. \ref{fig:Potenziale_ottico}) of three gaussian traps whose centers correspond to the vertexes of an equilateral triangle with edge $d$. The realization of such a potential is based on the projection of three dipole-trap beams from the vertical direction of our experimental apparatus which features a large numerical aperture objective with demonstrated high resolution capability \cite{gempel2019}. Being the wavelength 1064\,nm, we consider realistic values for the width of the three beams ($\sigma= 1,00\, \mu m$ ) and for the distance of their centers, $d=2\mu m$. For simplicity, we consider monochromatic tweezers, despite the large difference between sodium and potassium trap frequencies resulting from the different atomic polarizability. A bichromatic sheet of light on the horizontal direction or a time-averaged light potential provide confinement in the vertical direction and compensate the presence of a gravitational sag. 

\section*{The GPE solution in a real system}

In pursuing an actual experimental realization, one must interface the rather essential BH-like model (\ref{eq:BH}) described within the CVP (see effective potential (\ref{eq:V})) with experimentally-accessible parameters and measurable quantities. The bridge between theory and experiment is provided by a MF treatment of the problem. This approach also comes with a number of improvements to the CVP and, while confirming the fundamental results obtained therein, shines light on some possibly tricky aspects which were set aside by the pointlike approximation of potential wells and inherently present both in the BH model and in the CVP.

\begin{figure}[h!]
\centering
\includegraphics[width=1\linewidth]{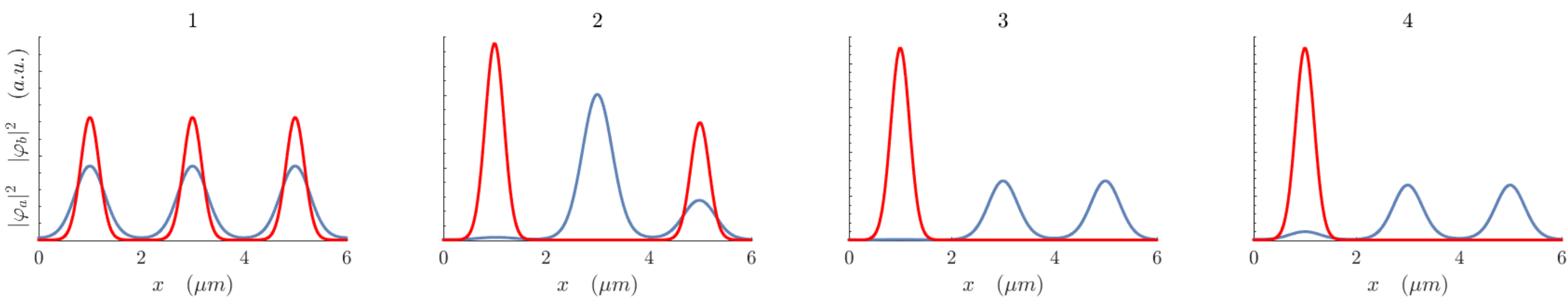}
\caption{Graphical representation of four qualitatively different eigensolutions of equations (\ref{eq:GPE1}) and (\ref{eq:GPE2}) which minimize Hamiltonian functional (\ref{eq:H_funcional}). Blue (red) line corresponds to $|\varphi_a|^2$ ($|\varphi_b|^2$). The horizontal axis constitutes the arc length associated to the perimeter of the equilateral triangle formed by the traps' centers (see yellow line in the right panel of Fig. \ref{fig:Potenziale_ottico}) and therefore it comes with periodic boundary conditions (see Methods for details concerning the numerical simulation). The specific model parameters that have been used in each simulation are given in the form $(N_a,\,N_b,\,g_a,\,g_b,\,g_{ab})$. 
Panel 1: ($200$, $600$, $52\times (4\pi \hbar^2 a_0)/m_a$, $8.1\times (4\pi \hbar^2 a_0)/m_b$, $10.2\times (2\pi \hbar^2 a_0)/\mu_{ab}$); 
Panel 2: ($200$, $600$, $52\times (4\pi \hbar^2 a_0)/m_a$, $7.6\times (4\pi \hbar^2 a_0)/m_b$, $28.4\times (2\pi \hbar^2 a_0)/\mu_{ab}$); 
Panel 3: ($200$, $220$, $52\times (4\pi \hbar^2 a_0)/m_a$, $7.8\times (4\pi \hbar^2 a_0)/m_b$, $48.5\times (2\pi \hbar^2 a_0)/\mu_{ab}$); 
Panel 4: ($200$, $220$, $52\times (4\pi \hbar^2 a_0)/m_a$, $7.7\times (4\pi \hbar^2 a_0)/m_b$, $23.0\times (2\pi \hbar^2 a_0)/\mu_{ab}$). All panels: the ring-trimer geometry is realized by means of three gaussian traps, whose centers lie on the vertexes of an equilateral triangle with edge $d=2 \mu m$, whose width is $\sigma=1 \mu m$ and whose depths are $P_a=4.44 \times 10^{-30}$ J = 322$\times k_B$\,nK for sodium and $P_b=1.16\times 10^{-29}$ J = 840$\times k_B$\,nK for potassium being $k_B$ the Boltzmann constant (see Methods). The energy of each of the four depicted solutions, together with their positions within the phase diagram spanned by effective parameters $g_{ab}/\sqrt{g_a g_b}$ and $N_b/N_a \sqrt{g_b/g_a}$ is shown in Fig. \ref{fig:Funzionali_energie}.}
\label{fig:Rappresentanti}
\end{figure}

In the MF picture, the system under analysis can be effectively modeled by two stationary GPEs
\begin{equation}
\label{eq:GPE1}
  -\frac{\hbar^2}{2m_a}\nabla^2 \varphi_a + g_a|\varphi_a|^2 \varphi_a+ g_{ab}|\varphi_b|^2 \varphi_a+V_{ext,a}\varphi_a =\mu_a \varphi_a
 \end{equation}
\begin{equation}
\label{eq:GPE2}
  -\frac{\hbar^2}{2m_b}\nabla^2 \varphi_b + g_b|\varphi_b|^2 \varphi_b+ g_{ab}|\varphi_a|^2 \varphi_b+V_{ext,b}\varphi_b =\mu_b \varphi_b
\end{equation}
with normalization conditions
\begin{equation}
\label{eq:Norm_cond_N}
  \int_{-\infty}^{+\infty} |\varphi_a|^2\, \mathrm{d}^3x= N_a, \qquad  \int_{-\infty}^{+\infty} |\varphi_b|^2\, \mathrm{d}^3x= N_b
\end{equation}
enforcing particle-number conservation in both atomic species. 
Formulas 
$$
g_a=\frac{4\pi \hbar^2 a_a}{m_a} =1.01\times 10^{-50} J \qquad \text{and} \qquad  g_b=\frac{4\pi \hbar^2 a_b}{m_b} \in[8.67,\,9.20]\times 10^{-52} J
$$
give the intra-species interactions and
$$
g_{ab}= \frac{2\pi \hbar^2 a_{ab}}{\mu_{ab}} \in[1.57,\,7.44]\times 10^{-51} J
$$
the inter-species coupling being $\mu_{ab}=(m_a^{-1}+m_b^{-1})^{-1}$ the reduced mass. These last two quantities can be tuned by means of the applied magnetic $B$ field, which, acting on scattering lengths $a_b$ and $a_{ab}$, therefore plays the role of the effective control parameter, capable of triggering the mixing-demixing transitions. Here we use $a_a=52\,a_0$, $a_b\in[7.6,8.1]\,a_0$ and $a_{ab}\in[10.2,48.5]\,a_0$. Term $V_{ext}(\vec{x})$ represents the optical potential corresponding to the ring-trimer geometry. One can generate (experimentally achievable) parameters' vectors ($N_a$, $N_b$, $g_a$, $g_b$, $g_{ab}$) both by tuning the magnetic field $B$ (the latter has an impact on ratios $g_{ab}/g_b$ and $g_b/g_a$) and by setting the boson population asymmetry $N_b/N_a$. On the basis of such sets of parameters, we determined (see Methods for details concerning the numerical simulation) the eigenfunctions of coupled, non-linear, eigenvalue problems (\ref{eq:GPE1}) and (\ref{eq:GPE2}), which minimize the Hamiltonian functional
$$
  \mathcal{H}[\varphi_a,\varphi_b] = \int \varphi_a^* \left[-\frac{\hbar^2}{2m_a}\nabla^2+ V_{ext,a}\right]\varphi_a\, \mathrm{d}^3 x +\frac{g_a}{2}\int |\varphi_a|^4 \, \mathrm{d}^3 x +
$$
\begin{equation}
\label{eq:H_funcional}
 + \int \varphi_b^* \left[-\frac{\hbar^2}{2m_b}\nabla^2+ V_{ext,b}\right]\varphi_b\, \mathrm{d}^3 x +\frac{g_b}{2}\int|\varphi_b|^4\, \mathrm{d}^3 x 
   + g_{ab}\int |\varphi_a|^2|\varphi_b|^2 \, \mathrm{d}^3 x,
\end{equation}
representing, in turn, the total energy of the system. Being qualitatively different, such minimum-energy solutions can be classified into four categories which indeed correspond to the already discussed four phases. In order to better appreciate the great agreement between the results obtained by means of the CVP and those obtained within the MF treatment, a representative of each of these categories is shown in Fig. \ref{fig:Rappresentanti}. 
With reference to such figure, the first panel represents the category of ground-states belonging to phase 1, the one exhibiting perfect mixing. The two condensates, in fact, clearly overlap in each of the three wells and both boson distributions share the same $Z_3$ symmetry of the system. Notice also that, correctly, sodium (depicted in blue), being lighter, is more delocalized with respect to potassium (depicted in red) which, being heavier, is more strongly confined. The second panel depicts a ground-state belonging to phase 2, where one has demixing in two wells and mixing in the remaining one. Notice, in fact, that the first (second) well includes just potassium (sodium), while the third well includes both. The third panel shows a fully-demixed ground state, i.e. a representative of phase 3: potassium occupies the first well while sodium spread in the second and in the third one. Eventually, the fourth panel shows a representative of phase 4, the one characterized by the clotting of potassium in a single well and by the presence of a non-zero fraction of sodium therein. 

Importantly, for all parameters' sets used in our numerical simulations, we observed that the phase separation mechanism is such that the two condensates redistribute \textit{among} the wells and not \textit{within} the wells. Nevertheless, since in a single harmonic trap different phase separation mechanisms (e.g. hemispheric-like or spheric-shell-like) can be triggered by different potential and interaction strengths\cite{Viktor}, we expect the interplay of these parameters to play an even more crucial role in multiple-trap systems. Due to its remarkable complexity, the analysis of this phenomenology will be discussed in a separate work.

\begin{figure}[h!]
\centering
\includegraphics[width=1\linewidth]{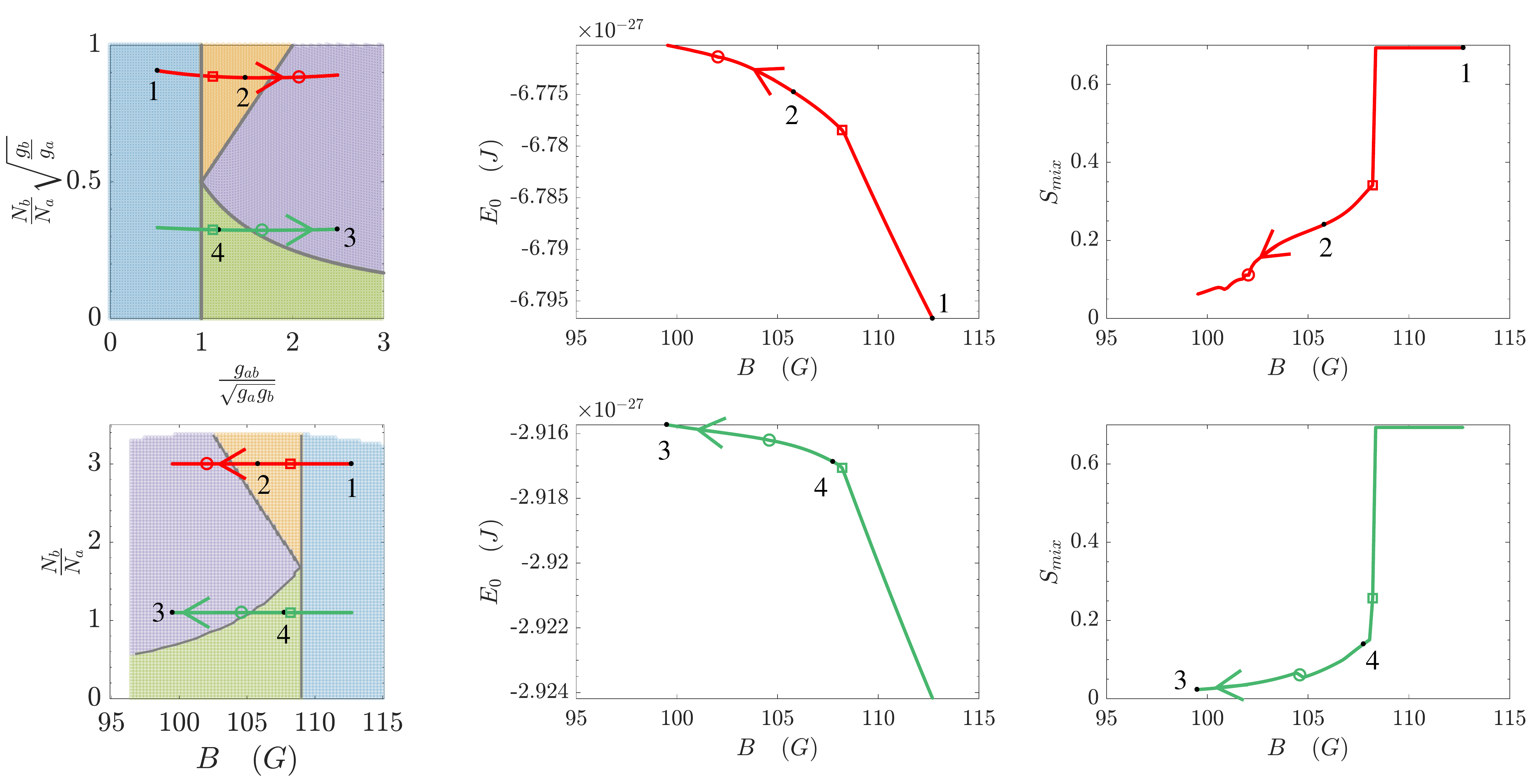}
\caption{Upper left panel: two pathways in the MF counterpart of the phase diagram shown in Fig. \ref{fig:Diagramma_di_fase}.  One can show, in fact, that parameter $\alpha$ and $\beta$ of Fig. \ref{fig:Diagramma_di_fase} play the role of $g_{ab}/\sqrt{g_a g_b}$ and $N_b/N_a \sqrt{g_b/g_a}$. 
Lower left panel: experimental counterpart of the phase diagram, the actual control parameters being the applied magnetic field $B$ and the ratio $N_b/N_a$. Notice that curved pathways in the upper left panel are here horizontal. Each pathway is obtained by sweeping the applied magnetic field $B$. The upper red (lower green) pathway is obtained for $N_a=200$ and $N_b=600$ ($N_b=220$).
Central column's panels: ground state energy computed along each pathway (more specifically, they are computed by plugging the solutions of equations (\ref{eq:GPE1}) and (\ref{eq:GPE2}) into functional (\ref{eq:H_funcional})). Rectangular (circular) markers indicate the points where the first (second) derivative of $E_0$ with respect to $B$ is discontinuous. Numbers 1-4 not only refer to the phase crossed by the pathway but specifically correspond to the minimum-energy solutions 1-4 illustrated in Fig. \ref{fig:Rappresentanti}. 
Right column's panels: entropy of mixing computed along each pathway. Normalized boson populations appearing in formula (\ref{eq:S_mix}) are computed by integrating the (square modulus of) the solutions of equations (\ref{eq:GPE1}) and (\ref{eq:GPE2}) in the three spatial domains associated to the potential wells. }
\label{fig:Funzionali_energie}
\end{figure}

It is worth noticing that, when trap parameters and interaction strengths are such that phase separation does occur via an among-well boson redistribution, the terms proportional to $g_a$, $g_b$ and $g_{ab}$ in energy functional (\ref{eq:H_funcional}) quite naturally exhibit the same structure of the terms proportional to $U_a$, $U_b$ and $W$ in effective potential (\ref{eq:V}). In this sense,  parameters $\alpha$ and $\beta$ (see formulas (\ref{eq:alfa_beta})) correspond to, respectively, quantities $g_{ab}/\sqrt{g_a g_b}$ and $N_b/N_a \sqrt{g_b/g_a}$, and therefore span the MF counterpart of the phase diagram of Fig. \ref{fig:Diagramma_di_fase}. To link the phase diagram to experimental accessible parameters, in Fig.\,\ref{fig:Funzionali_energie} we also plot the four phases by mapping $\beta$ into $N_b/N_a$ and $\alpha$ to the magnetic field $B$ thanks to the known values of scattering lengths shown in Fig.\,\ref{fig:Potenziale_ottico}. Accordingly, gray lines in the both left panels of Fig. \ref{fig:Funzionali_energie} play the role of the red boundaries separating the four phases in Fig. \ref{fig:Diagramma_di_fase}. 

As $g_b$ and $g_{ab}$ depend on the applied magnetic field $B$, by sweeping the latter, one can generate pathways through the phase diagram. Moreover, one can vertically translate them simply by tuning the ratio $N_b/N_a$. For many such pathways, practically constituting extended sets of n-tuples ($N_a$, $N_b$, $g_a$, $g_b$, $g_{ab}$), we found the solutions of equations (\ref{eq:GPE1}) and (\ref{eq:GPE2}) which minimize energy functional (\ref{eq:H_funcional}). Two examples of this analysis are shown in Fig. \ref{fig:Funzionali_energie}, where we plot the ground state energy $E_0:=\mathcal{H}[\varphi_{a,0},\varphi_{b,0}]$ as a function of the actual control parameter $B$ along two specific pathways through the phase diagram. In remarkable analogy with the discussion relevant to effective potential (\ref{eq:V_*}) within the CVP, it is clear that $E_0$ is a continuous function of $B$ and that its first derivative $\mathrm{d}E_0/\mathrm{d}B$ is discontinuous across transitions 1-2 and 1-4 (see rectangular markers in Fig. \ref{fig:Funzionali_energie}). Moreover, although not visually obvious, one can verify that the \textit{second} derivative $\mathrm{d}^2E_0/\mathrm{d}B^2$ is discontinuous at transitions 2-3 and 3-4 (see circular markers in Fig. \ref{fig:Funzionali_energie}). Correctly, one does not observe phase transitions exactly at grey boundaries shown in the phase diagrams. This is due to the presence of non-zero hopping terms which smooths and deforms the zero-hopping phase diagram. On the right panels of Fig.\,\ref{fig:Funzionali_energie} we plot the the mixing entropy $S_{mix}$ relative to the two paths previously considered and the presence of different mix/demix phases is confirmed. This is very promising in the outlook of an experimental realization as $S_{mix}$ represents the most obvious quantity obtainable from the measured atom number in each wells, irrespective of the observed permutation. We associate the presence of small fluctuations at small values of $S_{mix}$ in both panels to numerical resolution resulting from the strong demixing and the consequent very small atom number in the wells. One could consider these fluctuations as a statistical error on the simulations in that range of magnetic field strength.

Upon the associations $\alpha \leftrightarrow g_{ab}/\sqrt{g_a g_b}$ and $\beta \leftrightarrow N_b/N_a \sqrt{g_b/g_a}$, our MF analysis, based on realistic model parameters involving sodium and potassium atoms, has therefore led to the same phases and the same mixing-demixing transitions evidenced within the CVP and thus corroborates, both qualitatively and quantitatively, the predictions about the phase-separation mechanism obtained therein, offering, also, a viable path towards an actual experimental realization. 

\section*{Other experimental aspects}

To strengthen the feasibility of our proposal, we consider also a possible experimental sequence and problems related to the limited lifetime of the sample because of three-body losses.

The starting point of the experimental sequence will be the creation of the degenerate samples in a cross dipole trap \cite{PhysRevA.97.023623} at a magnetic field of about 154\,G and the following ramp of the magnetic field to the zero of the inter-species scattering length at 117\,G. This allows one to maximize the lifetime of the sample against three-body losses. The mixture can be adiabatically loaded into the trimer potential by rising the intensity of the tweezers and decreasing the intensity of the crossed dipole trap. A ramp of the magnetic-field strength to the target value follows, and the system is let mix/demix. After a fast ramp up of the tweezers intensity to stop hopping between the wells, the tweezers will be separated and the atom number of each species in the three wells will be detected with absorption imaging after a short time of flight. We expect that on each single realization, the system will lie in one of the possible cyclic permutations of the typical boson distributions. Post analysis will reveal the mixing/demixing depending on the applied magnetic field.

Once the interaction parameters have been fixed, mixing and demixing properties of the mixture do not depend on the absolute number of atoms but on their ratio through $\beta$. Envisioning an experimental realization, the atom number becomes critical as appearing with the third power into the three-body loss rate coefficient, which rules the lifetime of the mixture in the tweezers' wells. In the calculations we chose the order of magnitude of atom numbers in such a way that the lifetime exceeds the smallest of the two hopping time ($<20$ms, see Methods) of an order of magnitude. To this goal we calculate three-body densities from GPE simulations and consider three-body loss rate coefficients on the order of $10^{-41}$\,m$^6$/s, taken from experimental measurements performed the Hannover setup and compatible with literature values \cite{zaccantiEfimov}. 

\section*{Methods}

\paragraph{Numerical solution of stationary GPEs.}
Equations (\ref{eq:GPE1}) and (\ref{eq:GPE2}) constitute a non-linear, coupled, eigenvalue problem. Its solution is obtained by means of a numerical technique based on the damped oscillating particle method\cite{Magnus_PRE}. Preliminary, in order to lighten the computational burden, we enact a dimensional reduction (in the spirit of the scheme discussed in \cite{Salasnich_1,Salasnich_2}) to turn the original genuinely 3D problem into a 1D one. This is obtained by means of factorizations
$$
   \psi_a=\psi_a(x,y,z)=\psi_{a,x} (x) \psi_{a,y,z}(y,z) = \psi_{a,x}(x) \frac{1}{\sqrt{L_y}} \frac{1}{\sqrt{L_z}}
$$
$$
   \psi_b=\psi_b(x,y,z)=\psi_{b,x}(x) \psi_{b,y,z}(y,z)=\psi_{b,x}(x) \frac{1}{\sqrt{L_y}} \frac{1}{\sqrt{L_z}}
$$
where $\psi_a$ and $\psi_b$ are such that 
$$
  \varphi_a = \sqrt{N_a}\psi_a, \qquad  \varphi_b = \sqrt{N_b}\psi_b
$$
and are therefore normalized to unity, as a consequence of normalization conditions (\ref{eq:Norm_cond_N}). These substitutions lead to equations
\begin{equation}
\label{eq:Eig_prob_1}
    -  \frac{\partial^2 }{\partial x^2} 
 \psi_a + \gamma_a |\psi_a|^2\psi_a  + \gamma_{ab}|\psi_b|^2\psi_a +\tilde{V}_{ext,a} \psi_a  =\tilde{\mu}_a \psi_a 
\end{equation}
\begin{equation}
\label{eq:Eig_prob_2}
     - \frac{\partial^2 }{\partial x^2} \psi_b + \gamma_b |\psi_b|^2\psi_b  + \gamma_{ba}|\psi_a|^2\psi_b +\tilde{V}_{ext,b} \psi_b  =\tilde{\mu}_b \psi_b 
\end{equation}
(notice that wavefunctions labels "$a,x$" and "$b,x$" are understood) whose effective parameters are 
$$
   \gamma_a= \frac{2m_a g_a  N_a}{L_y L_z \hbar^2 } , \qquad \gamma_{ab}= \frac{2m_a g_{ab} N_b}{L_y L_z \hbar^2 },  \qquad \tilde{\mu}_a= \frac{2m_a}{\hbar^2}\mu_a,
$$
$$
   \gamma_b= \frac{2m_b g_b N_b}{L_y L_z \hbar^2 } , \qquad \gamma_{ba}= \frac{2m_b g_{ab} N_a}{L_y L_z \hbar^2 }, \qquad \tilde{\mu}_b= \frac{2m_b }{\hbar^2}\mu_b.
$$
where $L_y$ and $L_z$ are chosen in a such a way that $L_y\approx L_z\approx 2 \sigma \approx 2\mu m$.
Moreover, the optical-potential terms read:
$$
 \tilde{V}_{ext,a}= -\frac{2m_a }{\hbar^2} P_a  \sum_{i=1}^3  e^{-\frac{2(x-x_i)^2}{\sigma^2}}, \qquad  \tilde{V}_{ext,b}= -\frac{2m_b }{\hbar^2} P_b \sum_{i=1}^3  e^{-\frac{2(x-x_i)^2}{\sigma^2}} 
$$
where $x_i$'s are the positions of the centers of each gaussian trap, expressed in terms of the curvilinear abscissa describing the perimeter of the equilateral triangle formed by the $x_i$'s themselves (see yellow line in the right panel of Fig. \ref{fig:Potenziale_ottico}).  
Resorting to the damped oscillating particle method\cite{Magnus_PRE}, one can turn the search for the ground state of stationary non linear Schr\"{o}dinger equations (\ref{eq:Eig_prob_1}) and (\ref{eq:Eig_prob_2}) into the steady-state solution of the following damped oscillators:
\begin{equation}
\label{eq:Damped_oscillating_1}
  \frac{\partial^2\psi_a}{\partial \tau^2}+\eta\frac{\partial \psi_a}{\partial \tau} =  \frac{\partial^2 }{\partial x^2} \psi_a - \gamma_a |\psi_a|^2\psi_a -\gamma_{ab}|\psi_b|^2\psi_a -\tilde{V}_{ext,a}\psi_a + \tilde{\mu}_a \psi_a
\end{equation}
\begin{equation}
\label{eq:Damped_oscillating_2}
  \frac{\partial^2\psi_b}{\partial \tau^2}+\eta\frac{\partial \psi_b}{\partial \tau} =   \frac{\partial^2 }{\partial x^2} \psi_b - \gamma_b |\psi_b|^2\psi_b -\gamma_{ba}|\psi_a|^2\psi_b -\tilde{V}_{ext,b}\psi_b + \tilde{\mu}_b \psi_b
\end{equation}
where the possible differences between left and right members of equations (\ref{eq:Eig_prob_1}) and (\ref{eq:Eig_prob_2}) act as forcing terms, and where the presence of a non-zero damping $\eta$ guarantees the convergence to a steady-state solution. Upon recasting equations (\ref{eq:Damped_oscillating_1}) and (\ref{eq:Damped_oscillating_2}) into pairs of first-order equations by means of substitutions
$$
   q_a=\psi_a, \qquad p_a=\frac{\partial \psi_a}{\partial \tau} , \qquad    q_b=\psi_b, \qquad p_b=\frac{\partial \psi_b}{\partial \tau}, 
$$
one can solve the resulting dynamical system with standard finite difference methods\cite{Geometric_numerical}. More specifically, we made use of the leapfrog integrator \cite{Birdsall}, iterating it until equations (\ref{eq:Eig_prob_1}) and (\ref{eq:Eig_prob_2}) are satisfied up to \textit{relative} errors $<10^{-3}$ (in order to quantify the mismatch between left and right members, the 2-norm has been employed).   

\paragraph{The optical potential}

Acousto-optical modulators will generate the three beams and particular attention will be dedicated to calibrate the three wells in order to verify their being identical. To this goal, the use of digital mirror device (DMD) can be fruitful to compensate aberrations and defects in the potentials as already shown in other experiments\cite{Zupancic16}. We consider also the possibility to obtain trimer potentials based on triangular optical lattice. In this case the tighter confinement suggests that a much smaller atom number per site must be considered.

Species-specific trap frequencies are considered in the calculations as well as the different hopping rate coming from the different height of the inter-well barriers. These correspond to  $\omega_a\approx 2\pi\times5600$\,Hz,  $\omega_b\approx 2\pi\times6900\,Hz$. The depth of the wells is chosen to allow the hopping times \{1/$T_a$,1/ $T_b$\} to be smaller than 20ms for both species; estimated hopping rate are evaluated by calculating overlap integrals and one has \{$T_a,T_b$\} $\approx2\pi\times\{1000,50\}\,Hz$. Excessively large hopping rate would smooth out the phases previously discussed, while too small will not allow mix/demix in a time shorter than the mixture lifetime. Real hopping rate will be experimentally determined to overpass limitations coming from simple overlapping integral approximation.

\paragraph{Experimental stability}
One requires a large experimental stability of the three quantities involved in the mix/demixing phases. These are the atom number (preparation and detection), the trimer potential depth and the magnetic field strength. This last is below 30\,mG in the Hannover setup \cite{1810.00608} and allows a smooth tuning of the scattering length well within the resolution requested to observe the phase transition shown in Fig.\,\ref{fig:Funzionali_energie}. The optical-potential stability can be assured by advanced feedback techniques \cite{PhysRevA.92.021402} while the stability of the atom number by real-time analysis \cite{PhysRevLett.117.073604}.

\section*{Acknowledgements}
A.R. and V.P. wish to thank Magnus \"{O}gren and Edoardo Paganoni for their valuable help in the numerical methods. A.R. and A.Z wish to thank Luis Santos and Silke Ospelkaus for fruitful discussions. A.Z. gratefully acknowledges financial support from the European Research Council through ERC Starting Grant POLAR and from the Deutsche Forschungsgemeinschaft (DFG) through CRC 1227 (DQ-mat), project A03 and FOR2247, project E5.

\section*{Author contributions statement}
A.R. performed analytic and numerical calculations. A.Z. took care of the experimental aspects.  A.R., A.Z. and V.P. have analyzed the results and equally contributed in writing and reviewing the manuscript. V.P. supervised the work. 
\section*{Additional information}

\textbf{Competing interests:} The authors declare no competing interests. 

\end{document}